\documentclass[prd,a4paper,showpacs, nofootinbib, preprintnumbers,amsmath, multicol]{revtex4}

\usepackage{epsfig}
\usepackage{amssymb}
\usepackage{graphicx}

\usepackage{color}
\definecolor{rosso}{cmyk}{0,1,1,0.4}
\definecolor{rossos}{cmyk}{0,1,1,0.55}
\definecolor{rossoc}{cmyk}{0,1,1,0.2}
\definecolor{blu}{cmyk}{1,1,0,0.3}
\definecolor{blus}{cmyk}{1,1,0,0.6}
\definecolor{bluc}{cmyk}{1,1,0,0.1}
\definecolor{verde}{cmyk}{0.92,0,0.59,0.25}
\definecolor{verdec}{cmyk}{0.92,0,0.59,0.15}
\definecolor{verdes}{cmyk}{0.92,0,0.59,0.4}

\def\circa#1{\,\raise.3ex\hbox{$#1$\kern-.75em\lower1ex\hbox{$\sim$}}\,}

\makeatletter
\def\art{\@ifnextchar[{\eart}{\oart}}
\def\eart[#1]#2#3#4#5#6{{\rm #2}, {\em #3 \rm #4} {\rm (#6) #5 ({\em #1})}}
\def\hepart[#1]#2{{\rm #2, \em#1}}
\newcommand{\oart}[5]{{\rm #1}, {\em #2 \rm #3} {\rm (#5) #4}}

\newcommand{\beq}{\begin{equation}}
\newcommand{\eeq}{\end{equation}}
\newcommand{\bea}{\begin{eqnarray}}
\newcommand{\eea}{\end{eqnarray}}
\newcommand{\ba}{\begin{array}}
\newcommand{\ea}{\end{array}}
\newcommand{\bi}{\begin{itemize}}
\newcommand{\ei}{\end{itemize}}
\newcommand{\bn}{\begin{enumerate}}
\newcommand{\en}{\end{enumerate}}
\newcommand{\bc}{\begin{center}}
\newcommand{\ec}{\end{center}}

\newcommand{\gsim}{\lower.7ex\hbox{$\;\stackrel{\textstyle>}{\sim}\;$}}
\newcommand{\lsim}{\lower.7ex\hbox{$\;\stackrel{\textstyle<}{\sim}\;$}}

\begin{document}
\setcounter{page}{0}


\title{The Cold Spot as a Large Void: \\ Lensing Effect on CMB Two and Three Point Correlation Functions}

\author{ ~\\   Isabella Masina}  
\email{masina@fe.infn.it}
\affiliation{  {\it Dip.~di Fisica dell'Universit\`a degli Studi di Ferrara and INFN Sez.~di Ferrara, \\
Via Saragat 1, I-44100 Ferrara, Italy} }
\author{ \vskip -.3 cm and \\~   Alessio Notari}   
\email{alessio.notari@cern.ch}
\affiliation{  {\it  CERN, Theory Division, CH-1211 Geneva 23, Switzerland} }


\begin{abstract}
~~~\\

{\bf Abstract:}  The ``Cold Spot" in the CMB sky could be due to the presence of an anomalous huge spherical 
underdense region - a ``Void'' - of a few hundreds ${\rm Mpc}/h$ radius. Such a structure would have an impact 
on the CMB two-point (power spectrum) and three-point (bispectrum) correlation functions not only at low $\ell$, 
but also at high $\ell$ 
through Lensing, which is a unique signature of a Void. 
Modeling such an underdensity with an LTB metric, we show that for the power spectrum 
the effect should be visible already in the WMAP data only if the Void radius
is at least $L\gtrsim 1~{\rm Gpc}/h$, while it will be visible by the Planck satellite if $L\gtrsim 500~{\rm Mpc}/h$.
We also speculate that this could be linked to the high-$\ell$ detection of an hemispherical power 
asymmetry in the sky. Moreover, there should be non-zero correlations in the non-diagonal two-point function.
For the bispectrum, the effect becomes important for squeezed triangles with two very high $\ell$'s: 
this signal can be detected by Planck if the Void radius is at least $L \gtrsim 300~{\rm Mpc}/h$, while higher resolution experiments should be able to probe the entire parameter space.
We have also estimated the contamination of the primordial non-Gaussianity $f_{NL}$ due to this signal, 
which turns out to be negligible.

\end{abstract}


\baselineskip=14pt
\setcounter{page}{1}


\maketitle

\vskip .6cm
\section{Introduction}

One of the anomalies present in the  recent WMAP~\cite{WMAP} measurement of the 
Cosmic Microwave Background (CMB) is the so-called Cold Spot~\cite{ColdSpot1,ColdSpot2}: 
a large spherical region on an angular scale of about $10^\circ$ that appears to be anomalously cold.
The probability that such a pattern would derive from Gaussian primordial fluctuations is estimated to be 
about $1.8\%$, using the most conservative criteria~\cite{ColdSpot1,ColdSpot2}. While this could still be 
a statistical fluke, some authors have 
put forward the idea that it could be due instead to some object on the line-of-sight between us and the 
Last Scattering Surface (LSS)~\cite{Tomita,InoueSilk}. 

In a previous paper~\cite{MN} we have explored some observational consequences of the proposed idea
that the Cold Spot is due to a ``Void'' \cite{Tomita, InoueSilk} , {\it i.e.} to an anomalously 
large underdense region of some unknown origin. 
Traveling through a Void, photons are redshifted due to the fact that the gravitational potential is not 
exactly constant in time, the so-called Rees-Sciama (RS) effect~\cite{ReesSciama}. 
We have thus computed observational quantities associated to the RS effect, focusing in particular on 
the statistical properties of the CMB: the two-point (power spectrum) and  
three-point (bispectrum) correlation functions.

In the present paper we extend the analysis of~\cite{MN} by considering also the Lensing effect, namely 
the deflection that occurs to a CMB photon traveling through a Void. The interesting fact is that this 
deflection is only present if there is something on the line of sight, while it would be absent if the Spot 
is just a statistical 
fluke of the primordial large-scale fluctuations, thus representing a unique signature on the CMB maps, 
which can rule out or confirm the existence of such a Void. Moreover the Lensing effect is correlated 
with the RS effect, which can be seen in the some observables (such as the three-point correlation function \cite{MN}).

The paper is organized as follows.
In section II we consider the lensing produced by an underdense region and compute its profile and its 
decomposition in spherical harmonics.
In section III we compute the contribution to the two-point function.
In section IV we compute the contribution to the bispectrum, and evaluate the signal-to-noise ratio and the 
possible contamination on a primordial non-Gaussian signal.
In section V we draw our conclusions.
In appendix~\ref{AppISW} we discuss how our result change upon inclusion of a cosmological constant,
while in appendix~\ref{flatsky} we review the flat-sky approximation for the primordial bispectrum.

While this paper was in preparation, a similar approach has been taken by~\cite{spergel}, who have studied 
the lensing effect from a Void (and also from a texture), showing that high-resolution experiments, which 
focus on a small area of the sky, should be able to detect it. We briefly comment on this result in section III.

\vskip 1cm


\section{A Void in the line of sight: Rees-Sciama and Lensing effects}
\label{profiles}

As in~\cite{MN}, we consider the following cosmological configuration: an observer looks 
at the CMB through a spherical Void of radius $L$, located at comoving distance $D$ from us in the direction 
of the $\hat z$ axis. 
We assume that the Void does not intersect the LSS and that we are not inside it.
The observer receives from the LSS the CMB photons, whose fluctuations we assume to be adiabatic, 
nearly-scale invariant and Gaussian - as those generated {\it e.g.} by the usual inflationary mechanism. 
We will also assume that the location of the Void in the sky is not correlated at all with the primordial 
temperature fluctuations coming from inflation; this is true, for example, if such structure comes from a 
different process, such as nucleation of bubbles. 
For simplicity we disregard the effect of a cosmological constant. We show in appendix~\ref{AppISW} how the expressions can change, 
after inclusion of a cosmological constant.

We model the Void's inhomogeneous region via a spherically symmetric Lema\^itre-Tolman-Bondi (LTB) metric, 
matched to a Friedmann-Lema\^itre-Robertson-Walker (FLRW) flat model~\cite{MN}. 
Our density profile turns out to be the one of a "compensated" Void, {\it i.e.} the underdense central region 
is surrounded by a thinner external overdense region. This follows from the matching conditions, 
requiring that the Void does not distort the FLRW metric at large distance\footnote{For example, if the Void comes 
from a primordial bubble of true vacuum, this is a physically consistent requirement: in fact a bubble would have 
a thin wall with localized gradient energy, that compensates for the lower energy contained in the true vacuum in 
the interior region.}. Note that, due to the compensation, there will be no lensing effect on the photons 
which travel outside the LTB region. 
The Void is then characterized by the following quantities: i) the distance $D$ from us and its centre, 
ii) its radius $L$ (namely the radius of the inhomogeneous LTB region), iii) the shape of the Void density 
profile and iv) the amplitude of the density contrast, which we parameterize the value at the centre $\delta_0$ (which is defined to be negative). 
Note that, since the subtended angle $2 \theta_L$ is small, we have $\theta_L \approx L/D$.

Given our configuration, the observer detects one particular realization of the primordial Gaussian 
perturbations on the LSS {\it plus} the secondary effects due to this anomalous structure. 
As in~\cite{MN}, we want to give the theoretical prediction for the two-point and three-point correlation functions, 
in order to compare with the observations. We write the observed temperature fluctuation as the sum of three 
components:
\beq
\frac{\Delta T({\bf \hat n})}{T}~=~
\frac{\Delta T({\bf \hat n})}{T}^{(P)} +~
\frac{\Delta T({\bf \hat n})}{T}^{(RS)} +~
\frac{\Delta T({\bf \hat n})}{T}^{(L)} \, ,
\label{3temp}
\eeq
where (P) stands for primordial, (RS) for Rees-Sciama and (L) for Lensing. 
Each fluctuation is defined as:
\beq
\frac{\Delta T({\bf \hat n})}{T}^{(i)}\equiv ~\frac{T^{(i)}({\bf \hat n})-\bar{T}^{(i)}}{T} ~~~~,~~~~~~~~~~~
i=P,RS,L,
\eeq
with the bar representing the angular average over the sky and $T=\sum_i \bar{T}^{(i)}=2.73 K$.
The RS temperature fluctuation is smaller than the primordial one, 
but larger than the temperature fluctuation induced by Lensing.

\vskip 1 cm

\subsection{Rees-Sciama Temperature Profile}

We now briefly review \cite{MN} how to compute the shape for $\Delta T^{(RS)}/T$.
In~\cite{MN} we have modeled the Void in the line of sight as a perturbation of a FLRW metric with a 
Newtonian potential $\Phi$:
\beq
ds^2=a^2(\tau) \left[ -(1+2 \Phi)d\tau^2+(1-2 \Phi) dx^i dx^j \right]    \, ,
\eeq
where $\tau$ is the conformal time and where $x^i$ are dimensionless comoving coordinates. 
The potential $\Phi(r)$ is given as a function of the dimensionless comoving radial coordinate $r$ by:
\beq
\Phi(r)=-\frac{9 \, 3^{1/3}}{5~ (2 \pi)^{2/3}} \int_{r_L}^r k(\bar{r}) \bar{r} d\bar{r}  ~~~~~~,
~~~~~~r\leq r_L =  \frac{1}{2 (6 \pi)^{1/6}} L H_0~~, 
\label{potenziale}    
\eeq
while  $\Phi(r)$ vanishes for $r\ge r_L$. Here $H_0$ is the present Hubble constant $H_0= h/3000 \, ~{\rm Mpc}^{-1}$.
The arbitrary function $k(r)$ represents the local curvature and determines the shape of the density 
profile.  This approximation is valid as long as $k(r)$ is small. The only constraints on this function come from 
the smoothness of the density profile at the centre (which dictates $k'(0)=0$, where the prime denotes a derivative 
with respect to the $r$ coordinate) and from the requirement that the LTB patch matches to a flat FLRW universe 
($k'(L)=k(L)=0$). We have chosen arbitrarily the function $k(r)$ as follows:
\beq
k(r)=k_0 \left[1-\left(\frac{r}{r_L}\right)^{\alpha} \right]^2~~~~~~.
\label{kappa}    
\eeq
In the rest of this paper we focus on the case $\alpha=4$ but, as shown in~\cite{MN}, the results 
for other cases, {\it e.g.} $\alpha=2$, are not qualitatively different.

The RS temperature fluctuation is effectively described by two parameters: 
its amplitude at the centre of the Void, $A=\Delta T({\bf \hat z})^{(RS)}/T$, and its angular extension, 
{\it i.e.} the diameter of the cold region, $\sigma$. Clearly, $\sigma$ is smaller than $2 \theta_L$, 
which is the angle subtended by the full LTB region. The shape of the temperature profile (hence $\sigma$) 
is determined by the shape of the Void density profile. 
As for the amplitude $A$, we recall from~\cite{MN} that:
\beq
A \approx  0.5 ~\delta_0^2 ~(L H_0)^3 ~\left( 1-\frac{D H_0}{2}\right)= 0.5 ~\delta_0^2 ~\frac{(L H_0)^3}{\sqrt{1+z}}~~.
\label{eqARS}
\eeq
So, $A$ turns out to depend on the radius $L$ and the density contrast $\delta_0$, 
with a dependence on $D= L/\tan\theta_L$ which is mild unless the Void is close to the LSS. 
Note that the expression in eq.~(\ref{eqARS}) fully agrees with the one obtained in~\cite{InoueSilk}, 
except for our pre-factor which is three times larger, probably due to the different shape of the profile 
(see also appendix~\ref{AppISW}).  
We fix the numerical values of $A$ and $\sigma$ phenomenologically. Clearly, this leaves
a degeneracy in the choice of the physical parameters, since we have three of them 
($D$, $L$ and $\delta_0$) and only two observational constraints ($A$ and $\sigma$). 
In order to fix a range for the numerical values of the amplitude $A$, we proceed as in~\cite{MN} 
relying on the values given by~\cite{texture}, as follows: 
for the temperature at the centre we use the range $T=-(190\pm 80) \mu K$, 
which means $A= (7 \pm 3) \times 10^{-5}$;
for the angular size $\sigma$ of the cold region, we choose the particular but representative values $6^\circ$,
$10^\circ$ and $18^\circ$ (which correspond respectively to $\theta_L=7^\circ,11^\circ,20.5^\circ$).

\vskip 1cm

\subsection{Lensing Temperature Profile}

In this paper, our goal is to compute the contribution $\Delta T^{(L)}/T$ to the temperature fluctuation, 
due to the lensing of pre-existing primordial temperature fluctuations. 
The temperature fluctuation due to lensing is usually computed in a gradient expansion~\cite{Bernardeau, Lewis}, where the total fluctuation is given by
\beq
\frac{\Delta T}{T}({\bf \hat{n}})\sim \frac{\Delta T^{(P)}}{T}({\bf \hat{n}}) + \partial_i\frac{\Delta T^{(P)}}{T}({\bf \hat{n}}) \partial^i \Theta({\bf \hat{n}}) +  
\partial_i \partial_j \frac{\Delta T^{(P)}}{T}({\bf \hat{n}}) \partial^j \Theta ({\bf \hat{n}}) \partial^i \Theta ({\bf \hat{n}}) + \, ... \, .
\label{templensing}
\eeq
We call the correction terms $\Delta T^{(L)\, (1)}/T$ and $\Delta T^{(L) \, (2)}/T$, where the superscripts denote the perturbative orders.
In order to compute this, we need $\Theta$, the so-called Lensing potential, which is related to the gravitational 
potential $\Phi$ by the following line integral~\cite{Bernardeau,review, Lewis}:
\beq
\nabla_{\perp}\Theta= -2 \int_{\tau_{LSS}}^{\tau_{0}} d\tau \frac{\tau_{LSS}-\tau}{\tau_{LSS}} \nabla_{\perp} \Phi  \, ,
\label{lensing}
\eeq 
where $\tau_O$ and $\tau_{LSS}$ denote respectively the conformal time at the observer and at the LSS; 
$\nabla_{\perp}$ stands for a gradient in the direction transverse to the line of sight.
Now, given the gravitational potential, we can numerically compute the Lensing potential.
The way we proceed is that we approximate the trajectory as a straight line and we compute the integral, 
as a function of the observational angle $\theta$ 
(in spherical coordinates, having fixed the $\hat{z}$ axis towards the centre of the Void). 
The gravitational potential $\Phi$ is zero outside the LTB patch and positive at the centre (since we have a Void). 
Therefore its transverse gradient is negative, which means that the displacement vector $\nabla_{\perp}\Theta$ 
is positive (and vanishes at $\theta=0$ and $\theta=\theta_L$). 
Then, we integrate $\nabla_{\perp}\Theta(\theta)$ along the $\theta$ angle and we get 
$\Theta(\theta)$, which turns out to be negative. 

\begin{figure}[t!]\begin{center}\begin{tabular}{c}
\includegraphics[width=6.5cm]{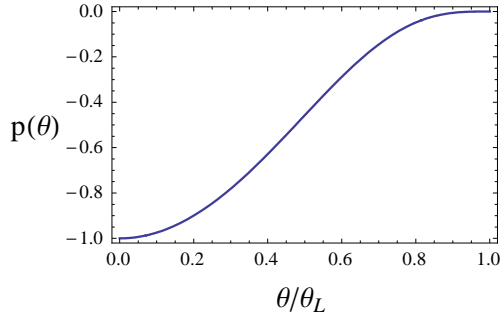} 
\end{tabular}\end{center}\vspace*{-0.5cm} 
\caption{Plot of the Lensing profile $p(\theta)$ as a function of $\theta/\theta_L$, 
for the curvature function given in eq.~(\ref{kappa}) and with $\alpha=4$.}
\label{profiliLENS}\vskip .7cm
\end{figure}

Now, the amplitude of the quantity $\nabla_{\perp} \Theta$ scales linearly with $\delta_0$ (since the effect 
is computed in the linear theory) and quadratically with the scale of the structure $L$: 
$
\nabla_{\perp}\Theta \propto \delta_0 \, (L H_0)^2 \, .
$
Removing the gradient, it turns out that $\Theta$ scales as $L^3$. The final equation for $\Theta$ can 
be written as\footnote{In the published version of this paper we neglected the dependence on $D$ coming from 
eq.~(\ref{lensing}). 
The inclusion of this dependence actually goes in the direction of strengthening the results as
compared to the published version.}:
\beq
\Theta(\theta)=\Theta_0 p(\theta)   ~~~~ , ~~~
\Theta_0 \approx \frac{1}{1.4} \, |\delta_0| \, (L H_0)^3 \frac{1}{D H_0}~~~,
\label{tetazero}
\eeq
where numerically $p(\theta)$ is given in fig.~\ref{profiliLENS}. 

For reader's convenience, we also give here an approximate polynomial interpolation of the profile:
\begin{equation}
p(\theta)= -1 - 0.257 \left( \frac{\theta}{\theta_L} \right) + 4.076 \left( \frac{\theta}{\theta_L} \right)^2 - 2.848 \left( \frac{\theta}{\theta_L} \right)^3~~~~,~~~\theta<\theta_L~~.
\end{equation}
Clearly $p(\theta)$ vanishes for $\theta \ge\theta_L$.

\begin{figure}[t!]\begin{center}\begin{tabular}{c}
\includegraphics[width=7cm]{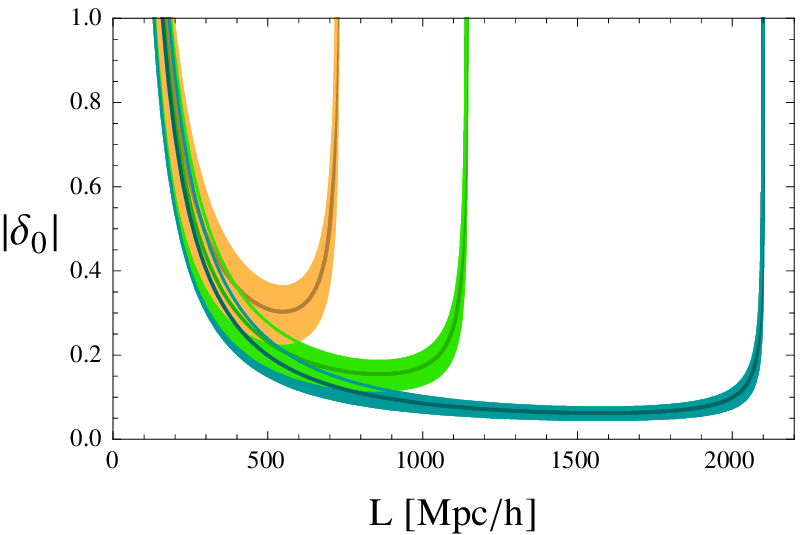}~~~~~~~~~~~~~~
\includegraphics[width=7.2cm]{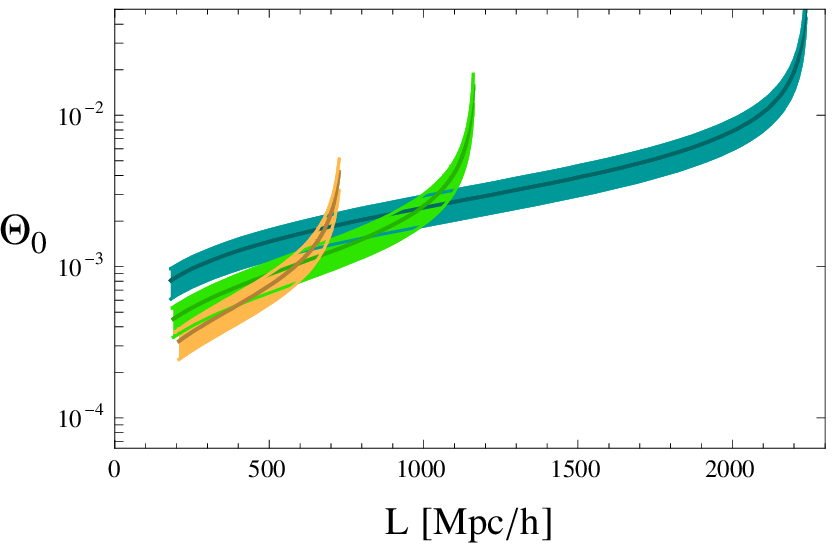} 
\end{tabular}\end{center}\vspace*{-0.5cm} 
\caption{Left: $|\delta_0|$ as a function of $L$ and for $\sigma=18^\circ, 10^\circ$ and $6^\circ$, 
from bottom to top. Right: Plot of $\Theta_0$ as a function of $L$. The curves corresponds, from left to right, 
to $\sigma=6^\circ,10^\circ,18^\circ$. In both plots,
the shaded regions are obtained by varying $A$ in the range $(7\pm 3) \times 10^{-5}$.}
\label{figL} \vskip .7cm
\end{figure}

Similarly to the RS temperature fluctuation, also the Lensing fluctuation, $\Delta T({\bf \hat n})^{(L)}/T$, 
is described by two parameters, namely those that characterize the Lensing potential $\Theta(\theta)$:   
its amplitude $\Theta_0$ at the centre of the Void and its angular extension, determined by $\theta_L$.
The amplitude $\Theta_0$ can be determined by choosing specific values for $L$ and $\delta_0$ in eq.~(\ref{tetazero}). 
Note, however, that the dependence of $\Theta_0$ on $L$ and $\delta_0$ is different
from the one of $A$: so it is possible for two Voids to have the same RS effect but different Lensing effects. 
Keeping fixed $A$ as indicated by the observations ($A= (7\pm 3) \times 10^{-5}$) 
and recalling that $D=L/\tan\theta_L$, 
we can calculate $\delta_0$ from eq.(\ref{eqARS}) by choosing $L$ and $\theta_L$: 
\beq
|\delta_0| \approx \sqrt{  \frac{2 A}{ 1-\frac{L H_0}{2 \tan \theta_L}} }  ~(L H_0)^{-3/2}~~~.
\label{eqdelta0}
\eeq
as shown in the left plot of fig.{\ref{figL}} for $\sigma=18^\circ, 10^\circ$ and $6^\circ$ (from bottom to top). 
Note that since $|\delta_0|\le 1 $, one finds a lower bound on $L$, with a very mild dependence on $\sigma$,
given by:
\beq
L \ge L_{min} \approx \frac{(2 A)^{1/3}}{H_0} \approx (150^{+30}_{-20}) ~\frac{{\rm Mpc}}{h}~.
\eeq
This minimum value of $L$ corresponds to the configuration in which the Void is closest to us.

We can put eq.(\ref{eqdelta0}) into eq.(\ref{tetazero}), obtaining
\beq
\Theta_0 \approx \left(\frac{A ~L H_0 \tan^2\theta_L}{1-\frac{L H_0}{2\tan\theta_L}} \right)^{1/2} ~~~.
\label{eqteta}
\eeq 
as shown in the right plot of fig. \ref{figL}. 
This figure shows that the dependence of $\Theta_0$ upon $\sigma$ turns out to be not too strong. 
For instance, the minimum value $\Theta_0=3\times 10^{-4}$ can be achieved with $L\approx 200$ Mpc/h for 
$\sigma=6^\circ$,  while the value $\Theta_0=10^{-3}$ can be achieved with 
$L=200-500$ Mpc/h for $\sigma=18^\circ$ and $L=500-800$ Mpc/h for $\sigma = 10^\circ$.

However, given a fixed value of $\theta_L$, both $\delta_0$ and $\Theta_0$ blow up at a certain $L$.
This maximum value of $L$ corresponds to the configuration in which the Void is very close to the LSS, 
without touching it.  
This fact can be understood via the following argument.
In a matter dominated Universe the angular size diameter distance of the LSS 
is $D_{A_{LSS}} \approx 2/(z_{LSS} H_0) \approx 5.45 \,  {\rm Mpc}/h$,  where $z_{LSS}=1100$. 
Hence, the maximum value of $L$ today is
\beq
L_{max} \approx  z_{LSS} ~\theta_L ~D_{A_{LSS}} ~~~,   
\eeq
where $\theta_L \approx L/D$.
For $\sigma=18^\circ$ ($\theta_L=20.5^\circ$), $\sigma=10^\circ$ ($\theta_L=11^\circ$) and
$\sigma=6^\circ$ ($\theta_L=7^\circ$) this corresponds respectively to $L_{max}$ about 
$2100  \,  {\rm Mpc}/h$, $1145  \,  {\rm Mpc}/h$, and $730  \,  {\rm Mpc}/h$.
These values are in agreement with fig.\ref{figL}. Close to the LSS, the amplitude $A$ 
is suppressed by the factor $1-D H_0/2$ of eq.(\ref{eqARS}).
Since by definition $|\delta_0|\leq 1$, there are values of $L$ such that the amplitude $A$ 
does not reach the value $7 \times 10^{-5}$.
For this reason there is a maximal value for $\Theta_0$, as shown in fig.~\ref{figL}.

On the other hand, when the Void is small, {\it i.e.} it is in the position closest to us, 
$\Theta_0$ reaches its minimum allowed value, which is about $3\times 10^{-4}$. Clearly, we find this minimum
because we are imposing $A$ to be in the range suggested by present Cold Spot observations, 
$A=(7\pm 3)\times 10^{-5}$:
had we imposed a smaller value of $A$, we would have obtained a smaller minimum value for $\Theta_0$.

Note also that assuming a different shape for the profile (see eq.~(\ref{kappa})) could lead to a different 
pre-factor in eq.~(\ref{eqARS}): instead of 0.5 we could have a pre-factor ranging from 0.05 to 1.8. 
This would just translate in a proper order-unity rescaling of eq.~(\ref{eqteta}) and fig.~\ref{figL}.


\vskip 1cm

\subsection{Decomposition in spherical harmonics} 

\begin{figure}[h!]\vskip .5cm\begin{center}\begin{tabular}{c}
\includegraphics[width=6.5cm]{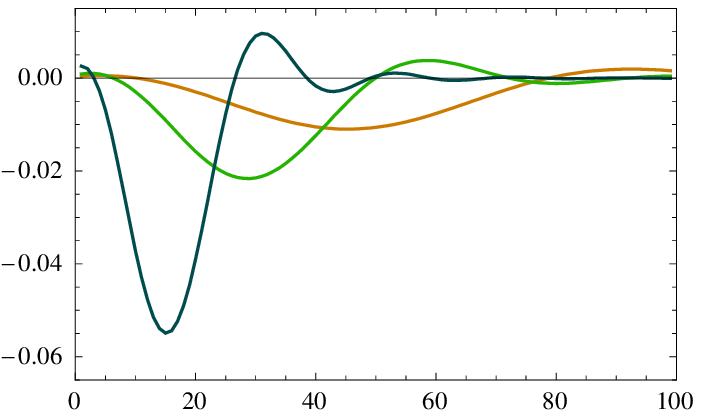} 
\put(-205,110){\Large $ \frac{a^{(RS)}_{\ell 0}}{A}$} \put(10,0){\Large $\ell$}~~~~~~~~~~~~~~~~
\includegraphics[width=6.5cm]{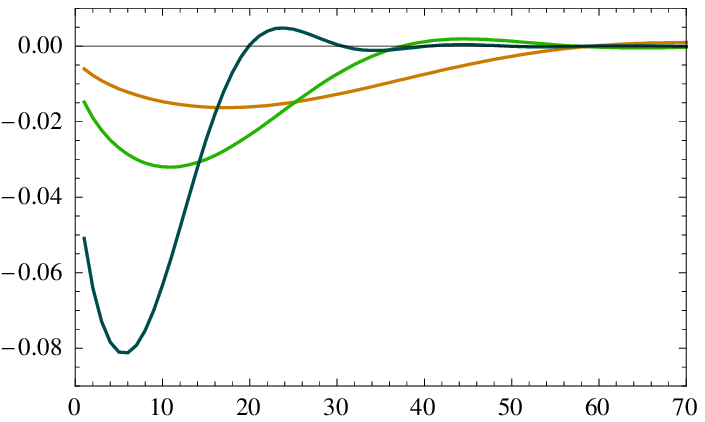} 
\put(-205,110){\Large $ \frac{b_{\ell 0}}{\Theta_0}$} \put(10,0){\Large $\ell$}
\end{tabular}\end{center}\vspace*{-0.5cm} 
\caption{Plot of $a^{(RS)}_{\ell 0}/A$ and $b_{\ell 0}/\Theta_0$, as a function of the multipole $\ell$
for $\sigma=18^\circ, 10^\circ, 6^\circ$ (from bottom to top). } 
\label{alm}\vskip .7cm
\end{figure}

Given the temperature anisotropy $\Delta T^{(i)}(\hat {\bf n})/T$ and the Lensing profile $\Theta(\hat {\bf n})$, 
we will need their spherical harmonic decompositions, defined respectively as:
\beq
a_{\ell m}^{(i)} \equiv \int d \hat {\bf n}~ \frac{\Delta T^{(i)}(\hat {\bf n})}{T}~Y^*_{\ell m}(\hat{\bf n})~~~~ 
 ,~~~~~~~~ b_{\ell m} \equiv\int d \hat {\bf n}~ \Theta(\hat{\bf n}) ~ Y^*_{\ell m}(\hat {\bf n})~~~~.  
\label{almblm}
\eeq
Since the RS temperature anisotropy and the Lensing profile are axially symmetric and since we have chosen 
the $\hat{z}$ axis to point towards the centre of the Void, the only non-vanishing $a^{(RS)}_{\ell m}$ and 
$b_{\ell m}$ are those with $m=0$ and which, in addition, are real.
In the left plot of fig.~\ref{alm} we show the ratio $a^{(RS)}_{\ell 0}/A$ as a function of the multipole $\ell$.
From bottom to top, the curves correspond to a Cold Spot with diameter $\sigma =18^\circ, 10^\circ$  and 
$6^\circ$ respectively (for more details see \cite{MN}).
In the right plot of fig.~\ref{alm} we show the corresponding $b_{\ell 0}/\Theta_0$ coefficients of the Lensing 
potential $\Theta(\hat {\bf n})$.

Given the $b_{\ell 0}$ coefficients in~(\ref{almblm}), we may compute the first order $a^{(L) \, (1)}_{\ell m}$ coefficients 
for the Lensing temperature profile $\Delta T^{(L) \, (1)}/T$, via the expression~\cite{review}:
\beq
a_{\ell m}^{(L) \, (1)}=\sum_{\ell', m', \ell''} G_{~\ell~~ \ell' \ell''}^{-m m' 0} (-1)^{m+m'}  
\frac{\ell'(\ell'+1)-\ell (\ell+1)+\ell''(\ell''+1)}{2} a^{(P)  *}_{\ell'-m'}  b_{\ell'' 0}  ~~ ,
\label{eqal1}
\eeq
where we have also used the fact that $b_{\ell'' \, m''} \propto \delta_{m'' 0}$, and that they are real numbers.  
We have introduced the Gaunt integrals, which are given in terms of the Wigner 3-j symbols~\cite{review} 
as follows:
\beq
 G_{\ell_1 ~\ell_2 ~\ell_3}^{m_1 m_2 m_3}\equiv \sqrt{\frac{(2 \ell_1+1)(2 \ell_2+1)(2 \ell_3+1)}{4 \pi}} 
 \left( \begin{matrix}  \ell_1 & \ell_2 & \ell_3 \cr 0 & 0 & 0 \end{matrix} \right)  
 \left( \begin{matrix}  \ell_1 & \ell_2 & \ell_3 \cr m_1 & m_2 & m_3 \end{matrix} \right) ~~ .
\eeq
The Gaunt integrals are non-zero only if the sum of the upper indices is zero, so we can set $m'=m$ in 
eq.(\ref{eqal1}) obtaining:
\beq
a_{\ell m}^{(L) \, (1)}=\sum_{\ell', \ell''} G_{~\ell~~ \ell' \ell''}^{-m m 0}  
\frac{\ell'(\ell'+1)-\ell (\ell+1)+\ell''(\ell''+1)}{2} a^{(P)  *}_{\ell'-m}  b_{\ell'' 0}    ~~ .
\label{almlensing}
\eeq
We recall that $\langle a^{(P)}_{\ell m}\rangle=0$, where the brackets stand for a statistical average 
over an {\it ensemble} of possible realizations of the Universe - or, equivalently, an average over many 
distant uncorrelated observers. Given eq.(\ref{almlensing}), it turns out that the same applies to the lensing 
coefficients, namely $\langle a^{(L) \, (1)}_{\ell m}\rangle=0$.

The expression for the second order coefficients $a_{\ell m}^{(L) \, (2)}$ has been derived {\it e.g.} in~\cite{Hu}, eq.~(46).

\vskip 1.5cm

\section{Two-point functions}

Given a temperature profile with its $a_{\ell m}$ coefficients one can compute the associated two-point 
correlation functions. 
In general, given {\it a single set} of $a_{\ell m}$ coefficients, the two-point correlation function (power spectrum) 
is defined via the $C_{\ell}$'s coefficients as
\beq
C_{\ell}\equiv\sum_{m=-\ell}^{\ell} \frac{|a_{\ell m}|^2}{2\ell+1}   \, .
\eeq
Note that this definition ensures that the $C_{\ell}$'s do not depend on the choice of the coordinate system on the 
sphere. 
Therefore, as in~\cite{MN}, we may keep our $\hat{z}$ axis aligned with the centre of the Void. In our case
$a_{\ell m}= a_{\ell m}^{(P)}+ a^{(RS)}_{\ell m}+ a^{(L)}_{\ell m}$ and, in order to face with the experimentally
observed values for $C_\ell$, we have to estimate the theoretical prediction for $\langle C_\ell \rangle$.

For a primordial Gaussian signal the two-point correlation functions are given by:
\beq
\langle a^{(P)}_{\ell_1 m_1} a^{(P) \, *}_{\ell_2 m_2} \rangle=\delta_{\ell_1 \ell_2} 
\delta_{m_1 m_2} \langle C^{(P)}_{\ell_1} \rangle   \ ,
\label{infpred}
\eeq
where the $\langle C^{(P)}_\ell \rangle$ are predicted by some mechanism ({\it e.g.} inflation) that can 
generate primordial Gaussian fluctuations. 
In~\cite{MN} we already computed the contribution to the power spectrum given by the RS effect, 
$\langle a^{(RS)} a^{(RS)*}\rangle$. Here we consider the Lensing term, which leads to additional 
contributions of the form $\langle a^{(P)} a^{(L)*}\rangle$ (PL contribution) 
and $\langle a^{(L)} a^{(L)*}\rangle $ (LL contribution).  The PL contribution actually contains  two terms: one is first order in the Lensing potential, $\langle a^{(P)} a^{(L) \, (1)*}\rangle$, and one is second order $\langle a^{(P)} a^{(L) \, (2)*}\rangle$.
It turns out that the latter term is of the same order as the $\langle a^{(L) \, (1)} a^{(L) \, (1)*}\rangle $ contribution and with opposite sign, leading to a cancellation\footnote{We thank A.~Challinor for pointing out this fact.}, as happens in Lensing from usual primordial Gaussian profiles~\cite{Lewis}.

Given the expression in eq.(\ref{almlensing}), we may compute the  first order contribution to the two-point correlation 
function due to the primordial and Lensing temperature fluctuations, as follows:
\begin{eqnarray}
\langle a^{(P)}_{\ell_1 m_1} a^{{(L)} \,(1) *}_{\ell_2 m_2} \rangle 
&=& \sum_{\ell', \ell''} G_{~\ell_2 ~~\ell'~ \ell'' }^{-m_2 m_2 0 }  
\frac{\ell'(\ell'+1)-\ell_2 (\ell_2+1)+\ell''(\ell''+1)}{2} (-1)^{m_2}
\langle a^{(P)}_{\ell_1 m_1} a^{(P)  *}_{\ell' m_2} \rangle b_{\ell'' 0} \nonumber \\ 
&=& \delta_{m_1 m_2} (-1)^{m_2} \langle C^{(P)}_{\ell_1} \rangle \sum_{\ell''} G_{~\ell_2 ~~\ell_1 ~\ell''}^{-m_2 m_2 0}  
\frac{\ell_1(\ell_1+1)-\ell_2 (\ell_2+1)+\ell''(\ell''+1)}{2}   b_{\ell'' 0}  \, .
\label{2pointPL}
\end{eqnarray}

The diagonal term leads to the following contribution to the $C_\ell$'s,
\beq
\langle C^{(PL)(1)}_{\ell} \rangle \equiv \sum_{m=-\ell}^{\ell} 
\frac{\langle a^{(P)}_{\ell m} a^{(L) \,(1) *}_{\ell m}\rangle }{2 \ell+1} ~~~ ,
\label{ClPL}
\eeq
which turns out to vanish, as one can see substituting the explicit expression for the Gaunt integral and using
a well known property of the Wigner 3-j symbols:
\beq
\langle C^{(PL)(1)}_{\ell} \rangle  =  \langle C^{(P)}_{\ell}\rangle \sum_{\ell''} \sqrt{\frac{2 \ell''+1}{4 \pi}} 
\left( \begin{matrix}  \ell & \ell & \ell'' \cr 0 & 0 & 0 \end{matrix} \right) \ell'' (\ell''+1)  b_{\ell'' 0} 
\underbrace{\sum_m (-1)^m \left( \begin{matrix}  \ell & \ell & \ell'' \cr -m & m & 0 \end{matrix}
 \right)}_{=\delta_{\ell''0}}=0~.
\eeq

The non-diagonal terms are quite interesting: in fact~(\ref{2pointPL}) leads to a correlation between different 
$\ell$'s. The correlations are small, but they are present also at high $\ell$'s. 
In fact any $\ell_1$ and $\ell_2$ will be correlated as long as $|\ell_1-\ell_2|\lesssim 60$, because the Gaunt 
integrals in~(\ref{2pointPL}) are nonzero if $|\ell_2-\ell_1|<\ell''$ and the $b_{\ell'' 0}$ coefficients are 
non-zero for $\ell\lesssim 60$ (for the values of $\sigma$ relevant for the Cold Spot, see fig.~\ref{alm}).
This signal could be of interest when looking at high-resolution experiments such as Planck, because 
it could reveal the existence of the large Void through the Lensing effect.
An interesting formalism to analyze these correlations in a rotational invariant way would be to introduce 
the so-called "Bipolar power spectrum"~\cite{souradeep}, which is a combination of the non-diagonal 
$a_{\ell m}$'s, which generalizes the $C_{\ell}$'s.
The Bipolar power spectrum is zero for Gaussian fields, but it has a contribution due to cosmic variance. 
It would be interesting to apply this analysis to our case and we postpone this to future work.
 
The pure Lensing contribution to the power spectrum, defined as
\beq
\langle C^{(LL)}_{\ell} \rangle 
\equiv \sum_{m=-\ell}^{\ell} \frac{\langle a^{(L)(1)}_{\ell m} a^{(L) (1)\, *}_{\ell m}\rangle }{2 \ell+1} 
~~~ ,
\label{ClLL}
\eeq
upon substitution with the expression in eq.(\ref{almlensing}), turns out to be:
\beq
\langle C^{(LL)}_{\ell} \rangle =  \sum_{\ell_1, \ell_2} \frac{2\ell_1 +1}{4\pi}
\left[ \frac{\ell_1(\ell_1+1) -\ell(\ell+1) + \ell_2 (\ell_2+1)}{2} \right]^2
\left( \begin{matrix}  \ell & \ell_1 & \ell_2 \cr 0 & 0 & 0 \end{matrix} \right)^2 (b_{\ell_2 0})^2~
\langle C_{\ell_1}^{(P)} \rangle   ~.
\label{corrLL}
\eeq
 Note that this expression agrees with the last term of eq.~(62) of \cite{Hu}, using the fact that the power spectrum of the Lensing potential in our case is $b_{\ell \, 0}^2/(2 \ell+1)$.
In addition to this term we have to add the term $\langle a^{(P)}_{\ell m} a^{(L) \, (2)} \rangle$, which is also given in eq.~(62) and~(63) of~\cite{Hu}, and which is equal to $- \ell (\ell+1)\, R \, \langle C_{\ell}^{(P)} \rangle$, where
\begin{equation}
R\equiv \frac{1}{8 \pi} \sum_{\ell} \ell (\ell+1) b_{\ell \, 0}^2 \, ,
\end{equation}
which can be easily computed and gives the result
\beq
R\approx 0.16 \, \Theta_0^2
\eeq
Therefore the final result for the correction $\Delta C_{\ell}$ to the power spectrum is given by~\cite{Lewis, Hu}:
\begin{equation}
\Delta C_{\ell}= - \ell (\ell+1) R \langle C_{\ell}^{(P)} \rangle + \langle C^{(LL)}_{\ell} \rangle \, .
\label{corrtotale}
\end{equation}
The first term is a multiplicative correction to the primordial power spectrum and the constant $R$ can be interpreted as the average of the square of the deflection angle~\cite{Lewis}. The perturbative treatment of Lensing is justified as long as this correction is small with respect to $\langle C_{\ell}^{(P)} \rangle$, that is $\ell (\ell+1) R \ll 1$. This means that the approximation is good until some $\ell_{\rm max}$, which is given in our case, by:
$$
\ell_{\rm max} \sim R^{-1/2} \sim 2.5/\Theta_0  \, .
$$
This is a good approximation for $\ell_{max} \lesssim 2000$, provided that $\Theta_0 \lesssim 10^{-3}$. In order to go to higher resolutions one should include the Lensing effect in a non-perturbative way, see {\it e.g.}~\cite{Lewis}.

\begin{figure}[t!]
\begin{center}\begin{tabular}{c}
\includegraphics[width=6.5cm]{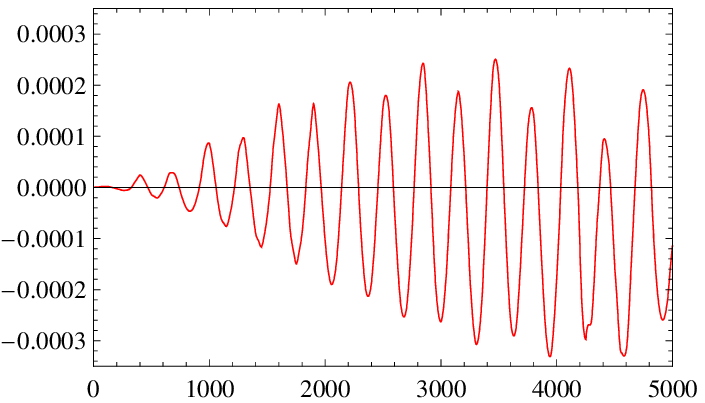} 
\put(-212,114){\Large $ \frac{\Delta C_{\ell} }{\langle {C}_{\ell}^{(P)} \rangle}$} 
\put(-5,-10){\Large $\ell$}~~
~~~~~~~~~~~~~~~~~~~
\includegraphics[width=6.5cm]{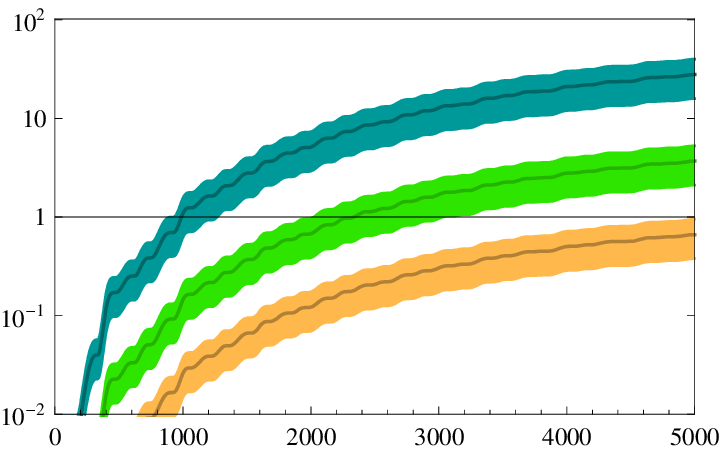} 
\put(-222,118){\Large $(S/N)_{LL}$} \put(-5,-10){\Large $\ell$}
\end{tabular}\end{center}\vspace*{-0.5cm} 
\caption{Left: Plot of $\Delta C_{\ell} /\langle {C}_{\ell}^{(P)} \rangle$ as a function of the multipole $\ell$
and with the minumum allowed value $\Theta_0=3\times 10^{-4}$. 
Right: Plot of $(S/N)_{LL}$ as a function of the multipole $\ell$. The shaded region 
is obtained by varying $A$ in the range $(7\pm 3)\times 10^{-5}$.
For the lower curve we have taken $L=200  \,  {\rm Mpc}/h$ and $\sigma=6^\circ$, which corresponds 
- see fig.\ref{figL} - to $\Theta_0= (3.1_{-.7}^{+.6})\times 10^{-4}$; 
for the middle curve we have taken $L=400  \,  {\rm Mpc}/h$ and $\sigma=10^\circ$, namely 
$\Theta_0=(7.4_{-1.8}^{+1.4})\times 10^{-4}$; 
for the upper curve we have taken $L=800 \,  {\rm Mpc}/h$ for $\sigma=18^\circ$, namely 
$\Theta_0=(2\pm .5)\times 10^{-3}$. 
Both curves are a good approximation only up to $\ell_{\rm max}\sim 2.5/\Theta_0$, while for higher $\ell$ there is an ${\cal O}(10\%)$ error, due to the fact that the gradient approximation breaks down~\cite{Lewis}.}
\label{SN2P}\vskip .7cm
\end{figure}

The second term in eq.~(\ref{corrtotale}) is a convolution: it can transfer power from large to small scales and in general can smooth some of the primordial peak structure.
The two terms are similar in magnitude but with opposite signs, and they cancel out with a precision of about $10^{-3}$, leaving an oscillatory correction.
The final result for the ratio $\Delta C_\ell /\langle C_\ell^{(P)}\rangle$ is shown in the left plot of fig.~\ref{SN2P} 
as a function of the multipole $\ell$, assuming a specific value for the amplitude of the 
Lensing potential, namely the minumum allowed one, $\Theta_0=3\times 10^{-4}$. Note that for different values of $\Theta_0$ one
has simply to multiply the results in fig.~\ref{SN2P} by the factor $\Theta_0^2/(9\times 10^{-8})$.

For an experiment whose sensitivity goes up to some $\ell_{max}$, the signal-to-noise ratio 
associated with the pure Lensing contribution is given by:
\begin{equation}
\left(\frac{S}{N}\right)_{LL}^2 = \sum_{2 \leq \ell \leq \ell_{max}} 
\frac{ \Delta C_{\ell}^2}{\sigma^2_{\ell}}  ~~~,
\label{signaltonoise}
\end{equation}

where $\sigma_{\ell}^2$ is the cosmic variance of the power spectrum: 
$\sigma^2_{\ell} \sim  2/(2\ell+1)~ \langle {C}_{\ell}^{(P)} \rangle^2$. 
In general $\sigma_{\ell}^2$ should contain also the experimental noise, but we have just cutoff 
the sum at the $\ell_{max}$ corresponding to the experimental sensitivity. 
The signal-to-noise ratio is displayed in the right plot of fig.~\ref{SN2P} for various Void configurations.
Note that $(S/N)_{LL}$ scales as $\Theta_0^2$. 

 Fig.~\ref{SN2P} shows that, if one assumes that the CMB Cold Spot is due to the RS effect of a Void with
central temperature fluctuation $A=(7\pm 3 )\times 10^{-5}$, then very high-resolution CMB experiments 
could detect $\Theta_0$: WMAP could see a signal only in the extreme case if $\Theta_0$ is
bigger than about $3\times 10^{-3}$ (which corresponds to a $\sigma$ of at least $10^{\circ}$ and $L\gtrsim 1000 {\rm Mpc}/h$, see fig.~\ref{figL}), 
while Planck could go down to $10^{-3}$, which corresponds to $L \gtrsim 500 {\rm Mpc}/h$ for $\sigma=10^\circ$. 
In order to be able to detect smaller Voids, one should look at the Cold Spot region 
with a higher resolution experiment (such as the planned Atacama Cosmology Telescope~\cite{Atacama} 
or the South Pole Telescope~\cite{Southpole}): {\it e.g.} going up to $\ell\sim 5000$ seems necessary 
to detect Voids with $L\sim 300 {\rm Mpc}/h$ and $\sigma=6^\circ$.


Moreover note that, clearly, if there is such an increase in the power spectrum due to a single Void, 
this would be localized in the Cold Spot region: therefore, masking this area the effect should be absent.
It is worth also mentioning that there could be other Voids of similar, or smaller, size in the sky, 
(for example~\cite{mystery} points out out that there might be other cold and hot spots of similar size
in the Southern hemisphere). This would lead to similar contributions to be added to the power spectrum, 
with different amplitudes.

An intriguing consequence of our analysis is that, if the Void is large enough, its Lensing signal could possibly be linked to the recent claimed detection 
of a hemispherical power asymmetry~\cite{hemispherical, hemispherical2} in the WMAP data, 
which extends also to high multipoles ($\ell=2-600$). 
While the low-$\ell$ power spectrum asymmetry 
could be linked to the existence of large structures (due to the RS effect, as discussed in~\cite{MN}), 
the fact that the signal is not limited to low multipoles could be explained by the lensing effect 
of such objects. Indeed, the direction of the center of the hemisphere with the largest power happens 
to be close to the location of the Cold Spot. The amplitude of the claimed effect~\cite{hemispherical2} 
is of order $\delta C_{\ell}/C_{\ell}\sim 10^{-3}$, which would be compatible with our effect 
(see fig.~\ref{SN2P}) if $\Theta_0$ is of order of a few times $10^{-3}$, which corresponds to 
$L\sim 1000 \, {\rm Mpc}/h$.
Therefore, it is observationally important to determine whether there is an increase in small-scale power, 
with the shape we predict in fig.~\ref{SN2P}, which could be linked to a huge Void, or possibly to several 
Voids and structures on large scales in the sky.

We may, finally, mention that the recent paper \cite{spergel} has proposed to estimate the amplitude of the 
lensing effect by looking directly at $\Delta T^{(L)}/T$, instead of looking at a quadratic estimator 
such as $\langle C^{(LL)}_{\ell}\rangle$. This leads to the consequence that the signal-to-noise ratio is linear 
in the $\langle C_{\ell}^{(LL)}\rangle$'s, instead of containing its squares, as happens for our eq.~(\ref{signaltonoise}). 
The maximal value of the lensing potential amplitude of~\cite{spergel} coincides with our 
$\Theta_0$ for $L=350 \,{\rm Mpc}/h$ and $\delta_0=0.3$, namely $\Theta_0=(7\pm 4) \times 10^{-4}$. 
Note however that~\cite{spergel} is considering a different geometry: a non-compensated cylinder 
of radius $150 \, {\rm Mpc}$ and transverse length $200 \, {\rm Mpc}$.
The signal $S_\ell$ in eq.~(13) of~\cite{spergel} is similar in structure to one of our terms, the $\langle C_{\ell}^{(LL)}\rangle$ in eq.~(\ref{corrLL}): in fact 
they turn out to have a very similar shape, although our $\langle C_{\ell}^{(LL)}\rangle$'s 
are smaller by a factor of order unity. However, the existence of an additional correction of the type $\langle a^{(P)} \, a^{(L) \, (2)} \rangle$ may affect, as in our case, the conclusions of~\cite{spergel}, since it might lead to a cancellation with a precision of order $10^{-3}$.

\vskip 1.5cm
\section{Bispectrum}

After having computed the $a_{\ell m}$ coefficients and the two-point correlation functions we may estimate 
the impact of a large Void on the bispectrum coefficients.
In~\cite{MN} we have considered the bispectrum due to the RS effect only, while in the present 
paper we focus on effects due to lensing.

The basic quantities are the $B_{\ell_1~\ell_2~\ell_3}^{m_1 m_2 m_3}$ coefficients, defined as:
\beq
B^{m_1 m_2 m_3}_{~\ell_1~ \ell_2 ~\ell_3} \equiv  ~ a_{\ell_1 m_1} ~ a_{\ell_2 m_2} ~ a_{\ell_3 m_3} \, .
\label{defBlm}
\eeq
These quantities are coordinate-dependent, so we define (in analogy with the $C_\ell$'s coefficients), 
the angularly averaged bispectrum 
\beq
B_{\ell_1 \ell_2 \ell_3} = \sum_{m_1,m_2,m_3} 
\left( \begin{array}{ccc} \ell_1 & \ell_2 & \ell_3 \\ m_1 & m_2 & m_3  \end{array} \right)
B^{m_1 m_2 m_3}_{~\ell_1 ~\ell_2 ~\ell_3}  \, .
\label{defBl}
\eeq
One can show, using the properties of the Wigner 3-j symbols~\cite{review}, that this combination does not depend 
on the chosen $z$-axis, so these are the quantities that we can use to make predictions.
For convenience we keep our $z$-axis along the direction of the centre of the Void. Then we perform the 
statistical averaging of the primordial perturbations.
As we have discussed in~\cite{MN}, using eqs.~(\ref{defBlm}) and~(\ref{almblm}) we get a sum of 27 terms. 
However, we want to give a prediction, {\it i.e.} compute statistical averages 
$\langle B_{\ell_1 \ell_2 \ell_3} \rangle$, and several terms have zero average. 
As we have said, a crucial assumption that we make is that the coefficients $a^{(RS)}_{\ell 0}$ are {\it not} 
stochastic quantities, which means that the location of the Void in the sky is not correlated at all 
with the primordial temperature fluctuations coming from inflation. 
Note that this is a conservative assumption: if there is some correlation, and barring cancellations, 
the non-gaussianity could be much more important since some terms would be non-zero. 
Under this assumption, we have shown in~\cite{MN}, that the two leading terms are 
\beq
\langle B^{(RS)}_{\ell_1 \ell_2 \ell_3}\rangle = \sum_{m_1,m_2,m_3} 
\left( \begin{array}{ccc} \ell_1 & \ell_2 & \ell_3 \\ m_1 & m_2 & m_3  \end{array} \right)
\langle a^{(RS)}_{\ell_1 m_1} a^{(RS)}_{\ell_2 m_2} a^{(RS)}_{\ell_3 m_3} \rangle 
\eeq
and
\beq
\langle B^{(PLRS)}_{\ell_1  \ell_2 \ell_3}\rangle
=  \sum_{m_1,m_2,m_3}  \left( \begin{matrix}  \ell_1 & \ell_2 & \ell_3 \cr m_1 & m_2 & m_3 \end{matrix} \right)
\langle a^{(P)}_{\ell_1 m_1} a^{(L)}_{l_2 m_2} a^{(RS)}_{l_3 m_3} \rangle + (5 \, \, {\rm permutations})   \, .
\eeq
The first one has been studied in detail in \cite{MN}. We now consider the second.

\vskip 1cm
\subsection{Signal-to-Noise ratio}

In this subsection we compute the contribution to $\langle B_{\ell_1 \ell_2 \ell_3} \rangle$ due to the coupling 
between the Primordial-Lensing and RS effects,  as follows:
\bea
\langle B^{m_1 m_2 m_3}_{\ell_1 ~\ell_2 ~\ell_3} \rangle^{(PLRS)}  
&=& f(1,2,3)+f(2,1,3)+f(1,3,2)+f(3,1,2)+f(2,3,1)+f(3,2,1) ~~~,\nonumber \\ \\
&~&~~~~~~~~~~~f(i,j,k)=  \langle a^{(P)}_{\ell_i m_i}   a^{(L)}_{\ell_j m_j}  a^{(RS)}_{\ell_k m_k} \rangle    \, .
\nonumber
\eea
Substituting the explicit expressions for the Lensing coefficients, eq.(\ref{almlensing}), we get:
\bea
f(i,j,k)&=& \sum_{\ell', \ell''} \langle a^{(P)}_{\ell_i m_i} a^{(P)  *}_{\ell'-m_j} \rangle  
G_{~\ell_j ~\ell' ~\ell''}^{-m_j m_j 0}  \frac{\ell'(\ell'+1)-\ell_j (\ell_j+1)+\ell''(\ell''+1)}{2}
 b_{\ell'' 0} a^{(RS)}_{\ell_k 0} \delta_{m_k 0} ~~~\nonumber \\
&=& \langle C_{\ell_i}^{(P)}\rangle \delta_{m_k 0} \delta_{m_i -m_j} a^{(RS)}_{\ell_k 0}
\sum_{\ell''} \frac{\ell_i(\ell_i+1)-\ell_j (\ell_j+1)+\ell''(\ell''+1)}{2} 
G_{\ell_j \ell_i \ell''}^{m_i -m_i 0} b_{\ell'' 0} ~~,
\label{eq-f}
\eea
where we have used the fact that $\langle a^{(P)}_{\ell_1 m_1} a^{(P)*}_{\ell'-m_2} \rangle 
= \langle C_{\ell_1}^{(P)} \rangle \delta _{\ell_1 \ell'} \delta_{m_1 -m_2}$.

The angularly averaged bispectrum is given by:
\beq
\langle B_{\ell_1 \ell_2 \ell_3}^{(PLRS)}\rangle = \sum_{m_1 m_2 m_3} 
\left( \begin{array}{ccc}\ell_1 & \ell_2 & \ell_3 \\m_1 & m_2 & m_3  \end{array} \right)
\langle B^{m_1 m_2 m_3}_{\ell_1 \ell_2 \ell_3} \rangle~~ ,
\eeq
and can be written as
\bea
\langle B_{\ell_1 \ell_2 \ell_3}^{(PLRS)}\rangle  &=& F(1,2,3) +F(3,1,2) +F(2,3,1) + (-1)^{\ell_1+\ell_2+\ell_3} 
(F(2,1,3) +F(1,3,2) +F(3,2,1))~~~,\nonumber \\ \label{eqB}\\
&~&~~~~~~~~~~~~~F(i,j,k) = \sum_{m_i,m_j,m_k} 
\left( \begin{matrix}  \ell_i & \ell_j & \ell_k \cr m_i & m_j & m_k \end{matrix} \right) f(i,j,k)~~~.\nonumber
\eea
Substituting the expression for the Gaunt integral in (\ref{eq-f}) we find
\bea
F(i,j,k) &=&  \langle C_{\ell_i}^{(P)} \rangle a^{(RS)}_{\ell_k 0} \sum_{\ell''} \sqrt{\frac{(2\ell_j+1) (2\ell_i+1) (2\ell''+1)}{4\pi}} 
\left( \begin{matrix}  \ell_j & \ell_i & \ell'' \cr 0 & 0 & 0 \end{matrix} \right)
\frac{\ell_i(\ell_i+1)-\ell_j (\ell_j+1)+\ell''(\ell''+1)}{2}  \nonumber \\
& & ~~~~~~~~~~~~~~~~~~~~~~~~~~~~~~~~~~~~\times b_{\ell'' 0} 
\underbrace{\sum_{m_i} \left( \begin{matrix}  \ell_i & \ell_j & \ell_k \cr m_i & -m_i & 0 \end{matrix} \right)
\left( \begin{matrix}  \ell_j & \ell_i & \ell'' \cr m_i & -m_i & 0 \end{matrix} 
\right)}_{= \frac{\delta_{\ell_k \ell''}}{2\ell_k+1}} 
\eea
so that, finally,
\beq
F(i,j,k)= \langle C_{\ell_i}^{(P)} \rangle a^{(RS)}_{\ell_k 0} \sqrt{ \frac{(2\ell_j+1)(2\ell_i+1)}{4 \pi (2\ell_k+1)} } 
\left( \begin{matrix}  \ell_j & \ell_i & \ell_k \cr 0 & 0 & 0 \end{matrix} \right)
\frac{\ell_i(\ell_i+1)-\ell_j (\ell_j+1)+\ell_k(\ell_k+1)}{2} b_{\ell_k 0} ~~.
\label{eqF}
\eeq

\begin{figure}[t!]\begin{center}\begin{tabular}{c}
\includegraphics[width=7.5cm]{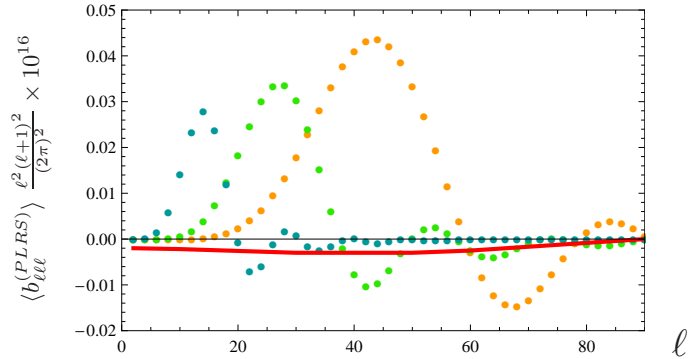} 
\put(-240,20){ \rotatebox{90}{$\langle b^{(PLRS)}_{\ell \ell \ell}\rangle 
~ \frac{\ell^2 (\ell+1)^2}{(2 \pi)^2} \times 10^{16}$}} 
\put(10,0){\Large $\ell$}
\end{tabular}\end{center}\vspace*{-0.5cm} 
\caption{The $\langle b_{\ell \ell \ell} ^{(PLRS)}\rangle$ coefficients - with $\ell$ even - due to 
Primordial-Lensing-RS for angular 
diameter $\sigma=18^\circ, 10^\circ, 6^\circ$ from left to right. 
We assumed a Void with $\Theta_0=3\times 10^{-4}$.
The solid line shows the primordial contribution with $f_{NL}=1$.}
\label{fig-blll}\vskip .5cm
\end{figure}

It is customary to define a reduced bispectrum $b_{\ell_1 \ell_2 \ell_3}$ via the following:
\beq
B_{\ell_1 \ell_2 \ell_3} = \sqrt{\frac{(2 \ell_1 +1)(2 \ell_2 +1)(2 \ell_3 +1)}{4 \pi}} 
\left( \begin{array}{ccc} \ell_1 & \ell_2 & \ell_3 \\ 0 & 0 & 0  \end{array} \right)~ b_{\ell_1 \ell_2 \ell_3} \, .
\label{reduced}
\eeq 
We may then easily compute, using~(\ref{reduced}), the $\langle b^{(PLRS)}_{\ell \ell \ell} \rangle$ coefficients:
\beq
\langle b_{\ell \ell \ell}^{(PLRS)} \rangle  =   \langle B_{\ell \ell \ell}^{(PLRS)}\rangle~ 
\frac{\sqrt{4 \pi}}{(2 \ell +1)^{3/2}} ~
\left( \begin{array}{ccc} \ell & \ell & \ell \\ 0 & 0 & 0  \end{array} \right)^{-1} \nonumber 
= \frac{3}{2} ~(1+(-1)^{3\ell})~  \frac{\ell (\ell+1)}{2 \ell +1} ~\langle C_{\ell}^{(P)} \rangle  
a^{(RS)}_{\ell 0} b_{\ell 0}~ .
\eeq
Note that $\langle b_{\ell \ell \ell}^{(PLRS)} \rangle =0$ if $\ell$ is odd.
The result for the $\langle b_{\ell \ell \ell}^{(PLRS)} \rangle $ with $\ell$ even 
is shown in fig.~\ref{fig-blll}, taking $\Theta_0=3\times 10^{-4}$ and, from left to right,
$\sigma=18^\circ, 10^\circ, 6^\circ$. 
The $\langle b^{(PLRS)}_{\ell \ell \ell}\rangle$ coefficients look comparable to the primordial non-Gaussianity 
with $f_{NL}$ of order unity (the solid line is obtained with $f_{NL}=1$) and they are non-zero for low $\ell$.

As can be seen by inspecting~(\ref{eqB}) and~(\ref{eqF}),
$\langle B_{\ell_1 \ell_2 \ell_3}^{(PLRS)}\rangle$ is non-zero provided that one of the $\ell$ indices is low, 
say $\ell \lesssim 60$ for $\sigma \ge 6^\circ$ (see fig.\ref{alm}). 
This means that, for an high resolution experiment as WMAP or Planck, which goes up to very high $\ell$, 
there are many non-zero coefficients, with $\ell_k \lesssim 60$ and arbitrary $\ell_i$ and $\ell_j$. 
The only constraint on $\ell_i$ and $\ell_j$ is given by the triangular condition of the Wigner 3-j symbols 
(which vanish unless $|\ell_i-\ell_j| \leq \ell_k \leq |\ell_i+\ell_j|$, for any $\ell_i$, $\ell_j$ and $\ell_k$).

We may indeed compute the Signal-to-Noise ($S/N$) ratio for the PLRS effect.
For a signal labeled by $i$, this is defined as \cite{review}:  
\begin{equation}
(S/N)_{i}=\frac{1}{\sqrt{F_{ii}^{-1}}}  ~~~,~~~~~
F_{ii}= \sum_{\substack{2 \leq \ell_1 \leq \ell_2 \leq \ell_3 \leq \ell_{max}}} 
\frac{ (B^{(i)}_{\ell_1 \ell_2 \ell_3})^2}{\sigma^2_{\ell_1 \ell_2 \ell_3}}   \, .
\label{Fiidef}
\end{equation}
Here $\sigma_{\ell_1 \ell_2 \ell_3}^2$ is the cosmic variance of the bispectrum
\beq
\sigma^2_{\ell_1 \ell_2 \ell_3}\sim \langle {\cal C}_{\ell_1} \rangle \langle {\cal C}_{\ell_2}\rangle \langle {\cal C}_{\ell_3}
\rangle \Delta_{\ell_1 \ell_2 \ell_3} \, ,
\eeq 
where $\Delta_{\ell_1 \ell_2 \ell_3}=1,2,$ or $6$ respectively if all $\ell$'s are different, if only two of them 
are equal or if they are all equal. 
The $\langle {\cal C}_\ell\rangle$'s are the sum of the CMB power spectrum plus the 
power spectrum of the noise of the detector. 
In general at some $\ell_{max}$ the noise becomes dominant, 
while below it $\langle {\cal C}_\ell\rangle$ is dominated by the primordial noise: 
$\langle {\cal C}_\ell \rangle\approx \langle C_\ell^{(P)} \rangle$.

\begin{figure}[t!]\begin{center}\begin{tabular}{c}
\includegraphics[width=13.5 cm]{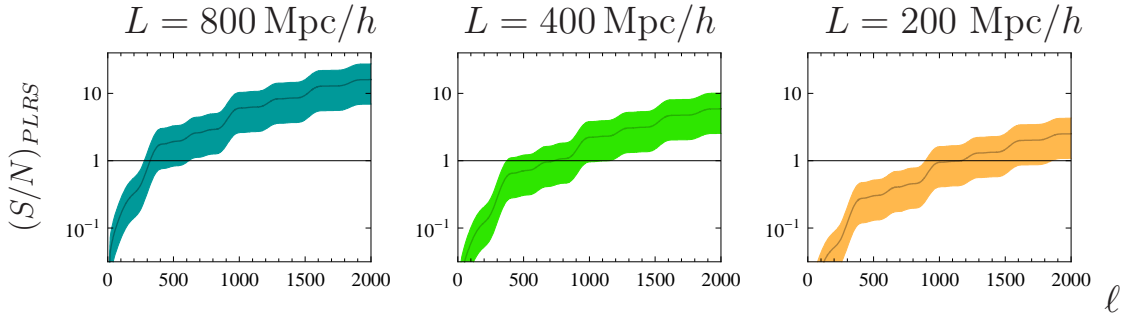} 
\put(-405,30){\large \rotatebox{90}{$(S/N)_{PLRS}$}} 
\put(5,0){\Large $\ell$}
\put(-362,105){\Large $L=800  \,  {\rm Mpc}/h$}\put(-235,105){\Large $L=400  \,  {\rm Mpc}/h$}\put(-100,105){\Large $L=200 ~{\rm Mpc}/h$}
\end{tabular}\end{center}\vspace*{-0.5cm} 
\caption{
The Signal-to-Noise ratio for the Primordial-Lensing-RS coupling.
The shaded regions are obtained by varying $A=(7\pm 3)\times 10^{-5}$.
For the right plot we have taken $L=200  \,  {\rm Mpc}/h$ and $\sigma=6^\circ$, which corresponds 
- see fig.\ref{figL} - to $\Theta_0= (3.1_{-.7}^{+.6})\times 10^{-4}$; 
for the middle plot we have taken $L=400  \,  {\rm Mpc}/h$ and $\sigma=10^\circ$, namely 
$\Theta_0=(7.4_{-1.8}^{+1.4})\times 10^{-4}$; 
for the left plot we have taken $L=800 \,  {\rm Mpc}/h$ for $\sigma=18^\circ$, namely 
$\Theta_0=(2\pm .5)\times 10^{-3}$.  }
\label{figuranuova}\vskip .5cm
\end{figure}

The result for $(S/N)_{PLRS}$ is shown in fig.~\ref{figuranuova}, assuming that the detector noise is negligible. 
Note that this ratio scales as the product $A \Theta_0$ and is nearly independent on $\sigma$.
Indeed, even though the coefficients $a_{l 0}^{(RS)}$ and $b_{l 0}$ display a dependence 
on the angle $\sigma$ (see fig.~\ref{alm}), we have found that the dependence of $(S/N)_{PLRS}$ on $\sigma$ 
is very weak for $6^{\circ} \leq \sigma \leq 18^{\circ}$.
Keeping $A$ fixed - the shaded region corresponds to letting $A$ vary in the range $(7\pm 3)\times 10^{-5}$ - 
the Lensing signal is clearly bigger the larger $\Theta_0$ (hence $L$) is. 
From left to right, the three panels in fig.~\ref{figuranuova} show $(S/N)_{PLRS}$ in the case that 
$L=800, 400, 200~ {\rm Mpc}/h$  (we have chosen for illustration various values of $\sigma$, corresponding
to $\Theta_0=(2\pm .5)\times 10^{-3}$, 
$\Theta_0=(7.4_{-1.8}^{+1.4})\times 10^{-4}$,
$\Theta_0= (3.1_{-.7}^{+.6})\times 10^{-4}$.
One can see that this signal is qualitatively different from the signal induced by the pure RS contribution
(see fig. 6 of ref. \cite{MN}) which becomes constant for $\ell_{max} \gtrsim 50$.
In fact, now the effect is tiny for small $\ell_{max}$, but it is sizable when $\ell_{max}$ is large. 
As a result, the total $(S/N)_{PLRS}$ {\it increases} with the sensitivity of the experiment, 
the physical reason being that any $\ell$ is lensed by the Void, also the small scale ones. 

As it can be seen from the plot, the ratio $(S/N)_{PLRS}$ summed over about 1900 multipoles  
(the maximal resolution for the Planck satellite) is significantly larger than unity, roughly if 
$L\gtrsim 300 ~{\rm Mpc}/h$. 
We stress therefore that this signature can uniquely discriminate the presence of a Void: in fact, 
if the Cold Spot is just a statistical fluke and if there is no Void (or a similar object) in the line of sight, 
the lensing effect would be absent.  It is also clear from the plot that going to slightly higher resolutions (as can be obtained by~\cite{Atacama} or ~\cite{Southpole}) will probe the entire parameter space.

\vskip 1cm
\subsection{Contamination of $f_{NL}$ measurements}
\label{deltafNLSECT}

We want here to find out what is the impact of having a hypothetical  huge Void in the line of sight 
on the measurement of the primordial non-gaussianity parameter $f_{NL}$.
It is customary to analyze the non-Gaussian signal via the quantity $f_{NL}$.
Its definition is given~\cite{review} for the so-called non-gaussianity of local type,
 parameterizing the primordial curvature perturbations $\phi(x)$ as follows
\beq
\phi(x)=\phi_L(x) + f_{NL} \left( \phi^2_L(x)-\langle \phi^2_L(x) \rangle \right)~~,
\eeq
where $\phi_L(x)$ is the linear Gaussian part of the perturbation.
Generically, given a physical model ({\it e.g.} slow-roll inflation), 
$f_{NL}$ is a function of the momenta, but it is usually assumed in quantitative data 
analyses to be just a number.
Single field minimally coupled slow-roll inflationary models predict generically very small values, 
$f_{NL} = {\cal O}(0.1)$~\cite{Maldacena,Acquaviva,review}, while other models may predict larger 
values~\cite{review}.

As discussed in \cite{MN}, the RS effect leads to a large correction to 
$\langle B_{\ell_1 \ell_2 \ell_3} \rangle $ for $2\leq \ell \leq 60$. 
The Lensing effect, instead, is much smaller at low-$\ell$, but it can contaminate 
the signal at large $\ell$, since it couples the low-$\ell$ of the RS effect with the high $\ell$ of the 
primordial signal. We now estimate the amount of such a contamination.

The primordial bispectrum coefficients can be written as~\cite{review}:
\beq
B^{prim}_{\ell_1 \ell_2 \ell_3}= f_{NL} \, \tilde{B}^{prim}_{\ell_1 \ell_2 \ell_3} \, ,
\eeq
where the $ \tilde{B}^{prim}_{\ell_1 \ell_2 \ell_3}$ have a specific form in terms of the primordial 
spectrum and the radiation transfer function~\cite{review}.

The impact on $f_{NL}$ due to the PLRS effect can be computed estimating the following 
ratio~\cite{review,cooray}:
\beq
\Delta f^{(PLRS)}_{NL}(\ell_{max})=
\frac{\sum_{2 \le \ell_1 \le \ell_2 \le \ell_3 \le \ell_{max}}\frac{B^{(PLRS)}_{\ell_1 \ell_2 \ell_3} \tilde{B}^{prim}_{\ell_1 \ell_2 \ell_3}}{\sigma^2_{\ell_1 \ell_2 \ell_3}} }
     {\sum_{2 \le \ell_1 \le \ell_2 \le \ell_3 \le \ell_{max}}\frac{(\tilde{B}^{prim}_{\ell_1 \ell_2 \ell_3})^2}{\sigma^2_{\ell_1 \ell_2 \ell_3}}} \, .
\label{deltafNLdef}
\eeq
In general this is computationally very demanding, since it would require to compute the primordial 
$\tilde{B}^{prim}_{\ell_1 \ell_2 \ell_3}$ up to very high multipoles. However, the effect due to the Void 
is dominated by squeezed triangles: since one side is given by the RS effect which goes up at most 
to $\ell\sim 80$, in order to compute the cross-correlation in the numerator of~eq.(\ref{deltafNLdef}) 
we only need the squeezed configurations also for the primordial effect.
We may give a rough estimate, using the so called flat-sky approximation, in which one of the 
$\ell$'s is much smaller than the other two. 
In this case we can write down a very simple expression for the 
$\tilde{b}^{prim}_{\ell_1 \ell_2 \ell_3}$ coefficients, following~\cite{Babich-Zaldarriaga}.
As discussed in Appendix~\ref{flatsky}, the result of this approximation turns out to be:
\beq
\tilde{b}^{prim}_{\ell_1 \ell_2 \ell_2} \approx -12 \, \langle C^{(P)}_{\ell_1}\rangle \langle C^{(P)}_{\ell_2}\rangle  ~~.
\label{BZ}
\eeq
The denominator of~eq.(\ref{deltafNLdef}) is instead a given quantity for a given experiment. 
In fact, the experimental noise in $\sigma^2_{\ell_1 \ell_2 \ell_3}$ at some high $\ell_{max}$ 
(dependent on the experiment) becomes so large that the multipoles $\ell> \ell_{max}$ 
do not contribute to the sum. The denominator in eq.(\ref{deltafNLdef}) is roughly equal to $ 10^{-8}\times 
\ell_{max}^2$,
as shown in \cite{review}.

Performing the sum we get a result which is of order of 
\begin{eqnarray} 
\Delta f^{(PLRS)}_{NL}& \approx & 10^{-5}\left( \frac{A}{7 \times 10^{-5}}\right) 
\left( \frac{\Theta_0}{10^{-4}}\right) \, ,
\end{eqnarray}
Using the maximal value for $A$ and $\Theta_0$, we can go up at most to $10^{-3}$.
The result has some $\ell_{max}$ dependence, but it is always of the order of $10^{-5}$, 
even for $\ell_{max}\sim 2000$.


\vskip 1.5cm
\section{Conclusions}

Motivated by the so-called Cold Spot in the WMAP data, we have studied in this paper 
the impact on CMB two and three point correlation functions of the presence of an anomalously 
large Void along the line of sight, whose existence could be at the origin of the Cold Spot. 
In particular, we have extended the analysis performed in~\cite{MN} by including the Lensing effect 
on the CMB primordial photons.

For the power spectrum, the Primordial-Lensing coupling vanishes exactly.
It leads, however, to non-zero correlations between different $\ell's$ in the off-diagonal 
two-point correlation functions. 
The Lensing-Lensing coupling, instead, gives a non-zero signal in the power spectrum, 
which turns out to have a Signal-to-Noise ratio of order unity already for WMAP sensitivity 
only if the Void radius $L$ is extremely large: bigger than about $1~{\rm Gpc}/h$ 
(since the amplitude of the Lensing effect increases with $L$). 
The Signal-to-Noise ratio turns out to be larger than unity for Planck sensitivity for $L\gtrsim 500 ~{\rm Mpc}/h$. 

In addition, we pointed out an intriguing consequence of our analysis: it could be possible that the existence 
of one (or several) large scale Void(s) in the sky appear not only as anomalies at low $\ell$, 
but also at high $\ell$, through the lensing effect.  In the case in which the Void is very large ({\it i.e.} $L \sim 1 ~{\rm Gpc}/h$),
this could be linked to the claimed detection of 
a hemispherical power asymmetry in the WMAP data~\cite{hemispherical}, 
which extends also to high multipoles ($\ell=2-600$) \cite{hemispherical2}.

For the bispectrum, there is a non-zero coupling between the Primordial fluctuations, the Rees-Sciama effect 
(present at low $\ell$) and the Lensing effect (present for any $\ell$). This would provide a distinctive signature in favor of a Large Void. In fact, its Signal-to-Noise exceeds unity 
for the Planck experiment if the radius of the Void is at least about $300 ~{\rm Mpc}/h$.  Higher resolution experiments will be able to detect this non-gaussian signal also for smaller Void radii, covering the entire parameter space.

Finally, we have studied the impact of such a structure on the determination of the primordial non-gaussianity, 
finding that the contamination is negligibly small.

\vskip 1.5cm
\acknowledgments We would like to thank Toni Riotto, Sudeep Das and Anthony Challinor for useful discussions and suggestions.

\vskip 1.5cm
\noindent {\bf Note added:}
In the published version of this paper we neglected the dependence of the Lensing Potential amplitude  
on the comoving distance of the Void, see eq. (\ref{tetazero}). 
The inclusion of this dependence actually goes in the direction of strengthening 
the results discussed in the published version.


\appendix

\vskip 1.5cm
\section{Inclusion of a cosmological constant}
\label{AppISW}

In this appendix we show how our results change by including the effect of a cosmological constant, 
which has the qualitative effect of increasing the RS effect because a net redshift is already present 
at the linear level in $\delta_0$~\cite{InoueSilk}. We remind the reader that in this case the effect 
usually goes under the name of Integrated Sachs-Wolfe effect (ISW). Therefore this additional ISW effect 
is especially relevant when $|\delta_0|$ is much smaller than unity, which is the case if the Void radius 
is very large, but still not close to touch the Last Scattering Surface (as can be seen in fig.~\ref{figL}).

\begin{figure}[t!]\begin{center}\begin{tabular}{c}
\includegraphics[width=7cm]{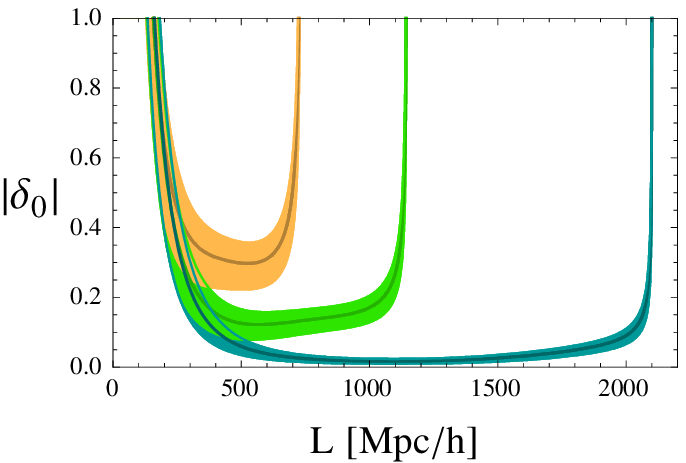}~~~~~~~~~~~~~~
\includegraphics[width=7cm]{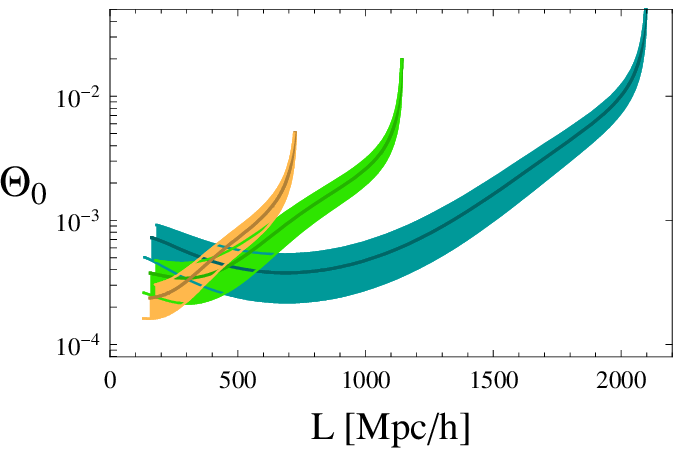} 
\end{tabular}\end{center}\vspace*{-0.5cm} 
\caption{Left: $|\delta_0|$ as a function of $L$, with $\Omega_M=0.27$ and for 
$\sigma=18^\circ, 10^\circ$ and $6^\circ$, from bottom to top. 
Right: Plot of $\Theta_0$ as a function of $L$, with $\Omega_M=0.27$. 
The curves corresponds, from top to bottom, to $\sigma=6^\circ,10^\circ,18^\circ$. 
In both plots, the shaded regions are obtained by varying $A$ in the range $(7\pm 3) \times 10^{-5}$.}
\label{figLambda} \vskip .5cm
\end{figure}

For this estimate we use the calculations performed by~\cite{InoueSilk}, with the {\it caveat} 
that they are performed in a different setup, 
with a different density profile, namely in the approximation in which the Void is uniformly underdense 
and it is compensated by an overdense thin shell. In the limit of zero cosmological constant our 
expression eq.~(\ref{eqARS}) fully agrees with the amplitude derived in~\cite{InoueSilk}, 
except for our pre-factor which is three times larger, probably due to the different shape of the profile.
We use the expression, multiplied by a factor of 3, given in~\cite{InoueSilk} for the amplitude $A$ of the 
(RS)+(ISW) effect in order to derive a relation between the density contrast $\delta_0$ at the centre of 
the Void and the radius of the Void (in such a way that $A=(7\pm 3 ) \times 10^{-5}$). This relation 
is shown in the left plot of fig.~\ref{figLambda}, choosing the matter density fraction $\Omega_M=0.27$. 
This can be directly compared with the left plot of fig.~\ref{figL}, for which $\Omega_M=1$.
Then, we can plug the new relation for $\delta_0$ into our eq.~(\ref{tetazero}), which allows us to 
reconstruct the value of $L$ given a measurement of the Lensing amplitude $\Theta_0$. We show the result 
for $\Omega_M=0.27$ in the right plot of fig.~\ref{figLambda}, which can then be compared with the right plot 
of fig.~\ref{figL} for which $\Omega_M=1$.

\begin{figure}[h!]\begin{center}\begin{tabular}{c}
\includegraphics[width=7cm]{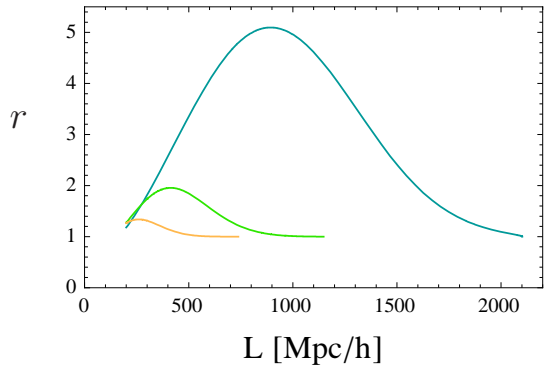}
\put(-200,90){\Large $r$} 
\end{tabular}\end{center}\vspace*{-0.5cm} 
\caption{Ratio $r$ defined as $\Theta_0$ with $\Omega_M=1$ over $\Theta_0$ with $\Omega_M=0.27$.
The curves correspond to $\sigma=6^\circ, 10^\circ$ and $18^\circ$, from bottom to top,
and to $A=7 \times 10^{-5}$.}
\label{figRTheta} \vskip .5cm
\end{figure}

For illustration, in fig.~\ref{figRTheta} we also plot the ratio of $\Theta_0$ obtained for $\Omega_M=1$ 
over the one with $\Omega_M=0.27$. As we have anticipated, the ratio is larger than $1$ in 
the regions of parameters where $\delta_0$ is much smaller than unity.
Anyhow, we stress again that this ratio is only a rough guide, given the fact that~\cite{InoueSilk} 
is using a different profile.

\vskip 1cm
\section{Flat-sky approximation}
\label{flatsky}

We re-derive here the approximation for the bispectrum~(\ref{BZ}), following~\cite{Babich-Zaldarriaga}. 
As a starting point we take the expression:
\begin{equation}\label{bprim}
   b^{prim}_{\ell_1 \ell_2 \ell_3} = 2 f_{NL} \int r^2 dr [\beta_{\ell_1}(r) \beta_{\ell_2}(r) \alpha_{\ell_3}(r) 
      + \beta_{\ell_1}(r) \alpha_{\ell_2}(r) \beta_{\ell_3}(r)  
      + \alpha_{\ell_1}(r) \beta_{\ell_2}(r) \beta_{\ell_3}(r)]~~,
\end{equation}
where
\begin{equation}
   \beta_{l}(r) = \frac{2}{\pi} \int k^2 dk P(k) j_l(kr) \Delta_{l}(k)~~,
\end{equation}
and 
\begin{equation}
   \alpha_{l}(r) = \frac{2}{\pi} \int k^2 dk j_l(kr) \Delta_{l}(k)~~.
\end{equation}
Here $j_l( kr)$ are the spherical Bessel functions, $P(k)$ is the
primordial power spectrum and $\Delta_l(k)$ is the 
radiation transfer function (which goes as 
$\Delta_l(k) \approx -j_l(kr_D)/3$ for large scales).
 Defining $\tau_O$ as the present day value of conformal time, 
$\tau_R$ as the value at decoupling and $r_D = \tau_0 - \tau_R$ as the comoving distance to
the surface of last scattering, 
the region of integration for $r$ is over the sound horizon (from $\tau_O$ to $\tau_0 - 2\tau_R$).
We can rewrite~(\ref{bprim}) as:
 \begin{equation}\label{physbisp}
      b^{prim}_{\ell_1 \ell_2 \ell_3} = \frac{16}{\pi^3} f_{NL} \int k_1^2dk_1 k_2^2dk_2 k_3^2dk_3 
        \Delta_{\ell_1}(k_1) \Delta_{\ell_2}(k_2) \Delta_{\ell_3}(k_3) 
        \mathcal{C}_{\ell_1 \ell_2 \ell_3}(k_1,k_2,k_3)
         [P(k_1)P(k_2) + cyc.]~~,
   \end{equation} 
   where we define $\mathcal{C}_{\ell_1 \ell_2 \ell_3}(k_1,k_2,k_3) 
   \equiv \int r^2 dr j_{\ell_1}(k_1 r)j_{\ell_2}(k_2 r) j_{\ell_3}(k_3 r)$.
In the limit of collapsed triangles, $\ell_1\ll \ell_2, \ell_3$, we can approximate
   \begin{equation}
      \mathcal{C}_{\ell_1 \ell_2 \ell_2}(k_1,k_2,k_3)  \sim j_{\ell_1}(k_1 r_D) 
      \int r^2 dr j_{\ell_2}(k_2 r) j_{\ell_2}(k_3 r)~~,
   \end{equation}
   and use the following property of the Bessel functions
   \begin{equation}
     \int r^2 dr j_{\ell_2}(k_2 r) j_{\ell_2}(k_3 r) \sim \frac{\pi}{2} \frac{\delta(k_2 - k_3)}{k^2_2}  ~~ .
   \end{equation}

 Substituting this result into~(\ref{physbisp}) we find
   \begin{equation}
       b^{prim}_{\ell_1 \ell_2 \ell_2} \sim \frac{8}{\pi^2} f_{NL} \int k_1^2 dk_1 k_2^2 dk_2 j_{\ell_1}(k_1 r_D) \Delta_{\ell_1}(k_1) 
       \Delta_{\ell_2}(k_2) \Delta_{\ell_2}(k_2) [P(k_1)P(k_2) + cyc.]~~.
   \end{equation}      
   This can be evaluated as
   \begin{equation}
       b_{\ell_1 \ell_2 \ell_2} \sim -12 f_{NL} \langle C^{(P)}_{\ell_1} \rangle \langle C^{(P)}_{\ell_2}\rangle~~,
   \end{equation}
   where we have used the fact that $\rangle C^{(P)}_{\ell}\rangle $ is given by:
\beq
\langle C^{(P)}_{\ell}\rangle=\frac{16 \pi^2}{(2 \pi)^3} \int dk k^2 |\Delta_{\ell}(k)| P(k) ~~ 
\eeq 
and where we have neglected the last term in the cyclic permutations, 
which is proportional to $P(k_2)^2$, since $P(k_1)\gg P(k_2)$.



\end{document}